\documentclass[%
 prb,
 amsmath,amssymb,floatfix,
 reprint,%
]{revtex4-2}

\usepackage{graphicx}
\usepackage{dcolumn}
\usepackage{bm}
\usepackage{amsmath}
\usepackage[utf8]{inputenc}
\usepackage[T1]{fontenc}
\usepackage{etoolbox}
\usepackage{color}

\makeatletter
\def\@email#1#2{%
 \endgroup
 \patchcmd{\titleblock@produce}
  {\frontmatter@RRAPformat}
  {\frontmatter@RRAPformat{\produce@RRAP{*#1\href{mailto:#2}{#2}}}\frontmatter@RRAPformat}
  {}{}
}
\makeatother
\begin{document}
\title{Surface magnon spectra of nodal loop semimetals}
\author{Assem Alassaf}
\affiliation{
 Department of Physics of Complex Systems, ELTE E{\" o}tv{\" o}s Loránd University, P{\'a}zmány P{\'e}ter s{\' e}t{\' a}ny 1/A, 1117 Budapest, Hungary}
\email{assem.al.assaf.abd.alrham@ttk.elte.hu}

\author{János Koltai}
\affiliation{
 Department of Biological Physics, ELTE E{\" o}tv{\" o}s Loránd University, P{\'a}zmány P{\'e}ter s{\' e}t{\' a}ny 1/A, 1117 Budapest, Hungary
}

\author{László Oroszlány}
\affiliation{
 Department of Physics of Complex Systems, ELTE E{\" o}tv{\" o}s Loránd University, P{\'a}zmány P{\'e}ter s{\' e}t{\' a}ny 1/A, 1117 Budapest, Hungary; MTA-BME Lendület Topology and Correlation Research Group, Budafoki {\'u}t 8., H-1111 Budapest, Hungary}%

\date{\today}

\begin{abstract}
In this paper we establish a connection between the bulk topological structure and the magnetic properties of drumhead surface states of nodal loop semimetals. We identify the magnetic characteristics of the surface states and compute the system's magnon spectrum by treating electron-electron interactions on a mean-field level. We draw attention to a subtle connection between a Lifshitz-like transition of the surface states driven by mechanical distortions and the magnetic characteristics of the system. Our findings may be experimentally verified e.g. by spin polarized  electron energy loss spectroscopy of nodal semimetal surfaces. 
\end{abstract}

\maketitle

\section{\label{sec:intro}Introduction}
Due to their unique electronic properties and potential applications in numerous fields, topological materials have attracted significant attention \cite{hasan2010colloquium, qi2011topological,fang2016topological,yan2017topological,armitage_RevModPhys.90.015001_topological}. These materials possess nontrivial topological properties, that in some cases need be protected by symmetries, resulting in the existence of robust surface or edge states. Topological semimetals are a class of topological materials that have been extensively studied in recent years \cite{wan2011topological, burkov2011weyl}. Weyl and nodal semimetals are two types of topological semimetals that possess distinct surface states. Weyl semimetals are distinguished by the presence of Weyl nodes in the bulk band structure, resulting in Fermi arcs on the surface \cite{wan2011topological, lv2015experimental}. These Fermi arcs connect the Weyl node projections and exhibit a variety of fascinating transport properties. In contrast, the drumhead states on the surfaces of nodal semimetals are dispersionless states associated to the surface projection of the nodal line structure. Due to their small kinetic energy these drumhead states are susceptible to interactions and thus they can be an ideal platform for superconductivity\cite{Rahul_bulk_supra_PhysRevB.93.020506,Rahul_surface_supra_PhysRevB.95.060506} or emergent surface magnetism \cite{burkov2011weyl, fang2015topological}.
Rhombohedral graphite is a prime example of such a material whose interaction induced magnetic properties have already been studied theoretically \cite{Campetella_PhysRevB.101.165437,ED_paper2021exchange} and observed experimentally \cite{zhou2021half,hagymasi2022observation}.

In this paper we investigate, through a simple model, the surface magnon spectrum of nodal loop semimetals. In the next section we introduce our model and describe the connection of the bulk nodal loop and drumhead surface states.  Treating  electron-electron interaction on a mean field level we obtain the magnetic properties of the surface states. Mapping to an isotropic Heisenberg spin model, we calculate the magnon spectrum of the system. We highlight a nuanced connection between the connectivity of the topological flat band and the magnon energies. Our findings should be relevant for experimental characterization of topological flat bands arising in nodal semimetals, specially when the flat bands extend over a considerable portion of the projected Brillouin zone, for example as those in Ca$_3$P$_2$ \cite{ca3p2_experiment,ca3p2theory_PhysRevB.93.205132}.

\begin{figure}
{\centering\includegraphics[width=\columnwidth]{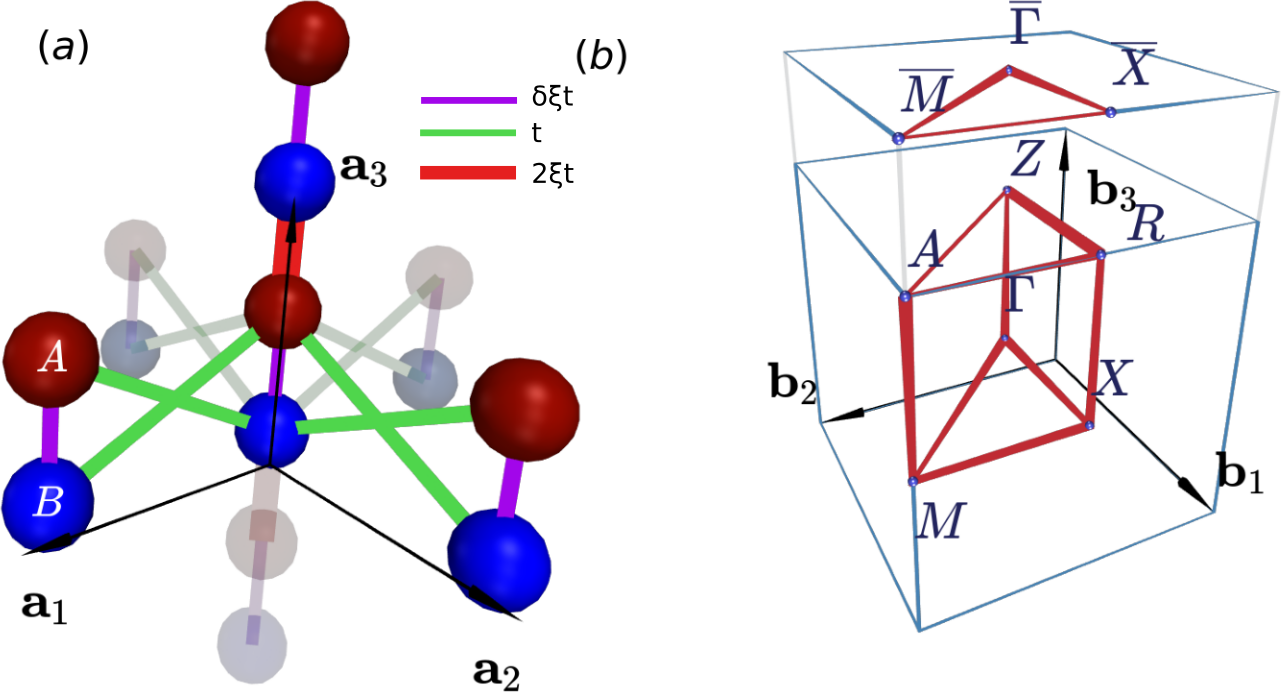}}
\caption{\label{fig:real_and_BZ} Real space structure (a) and high symmetry points in the full and projected  Brillouin zone (b) of the considered model. 
}
\end{figure}

\section{\label{sec:model}The Model}
In this section we introduce the investigated model and describe the real space structure and momentum space spectrum. The presence of a nodal loop, which is a closed curve in momentum space, distinguishes our model. As we show, the shape of the nodal loop and the flat surface states stabilized by its presence can be controlled by a parameter which corresponds to mechanical distortion in an experimental setting. 

\subsection{\label{sec:real}Real space structure}
We consider a three dimensional cubic system, spanned by the lattice vectors $\mathbf{a}_i$ with two sublattices (A and B). The real space structure is depicted in Fig.~\ref{fig:real_and_BZ} (a). We take a single spinfull orbital degree of freedom on each site into account. Electrons are allowed to hop from one site to the other without breaking of the sublattice symmetry characterized by the real space Hamiltonian:
\begin{equation}\label{eq:H0}
\begin{aligned}
\hat{H}_0 = \sum_{\mathbf{r},s} &\delta \xi t {\hat{a}_{\mathbf{r},s}^\dagger \hat{b}_{\mathbf{r},s} + t {\hat{a}_{\mathbf{r},s}^\dagger \hat{b}_{\mathbf{r}+\mathbf{a_1},s}}} \\
&+ t {\hat{a}_{\mathbf{r},s}^\dagger \hat{b}_{\mathbf{r}+\mathbf{a_2},s}} + 2\xi t  {\hat{a}_{\mathbf{r},s}^\dagger \hat{b}_{\mathbf{r}+\mathbf{a_3},s}} +\text{h.c. ,}
\end{aligned}
\end{equation}
where $\mathbf{r}$ represents a unit cell of the system, while $s$ is the spin degree of freedom. The annihilation operators $\hat{a}_{\mathbf{r},s}$ and $\hat{b}_{\mathbf{r},s}$ act on the appropriate sublattice and spin degree of freedom. The hopping amplitude $t$ controls the strength of electron movement between neighboring lattice sites and serves as the unit of energy for our model. The sublattice symmetry is the fundamental symmetry of the system which allows for the emergence of the nodal loop. 

There are two more important dimensionless parameters in the considered system. The parameter $\delta$ serves as an internal parameter which mimics experimentally hard to control properties of the system, such as particular matrix elements of the Hamiltonian related to hopping from one orbital to the other, while $\xi$, multiplying all hopping amplitudes in the $z$ - direction, captures effects of applying mechanical pressure on the system. As we shell see below, both of these parameters have a significant impact on the electronic structure and magnetic properties of the system, as they both control the shape of the nodal loop and the associated drumhead surface states. 

\begin{figure*}[ht!]
\begin{minipage}[b][0.5\textheight]{\textwidth}
\includegraphics[width=\textwidth]{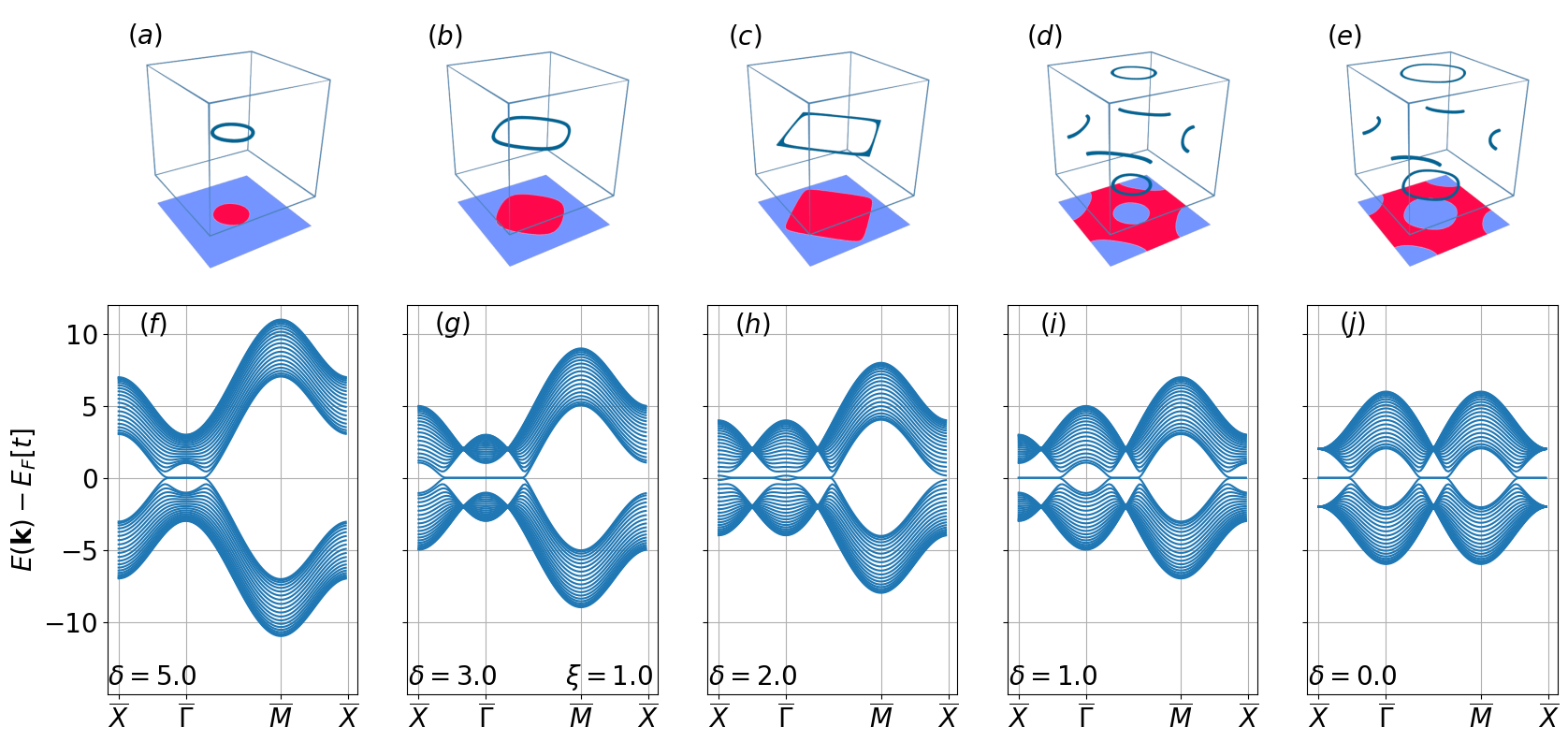}
\caption{\label{fig:band_structure_all}
The geometry of the nodal loop in the three dimensional Brillouin zone and the winding number map in the projected Brillouin zone for various values of $\delta$ (a)-(e). Purple shades in the lower plane correspond to winding number $\nu=0$ while red signals $\nu=1$. 
The band structure of a finite slab with thickness of 20 unit cells evaluated on a high symmetry path in the projected Brillouin zone for the same values of $\delta$ as above (f)-(j). In all cases $\xi=1.0$. 
}
\end{minipage}
\vspace{-10mm}
\begin{minipage}[b][0.5\textheight]{\textwidth}
\includegraphics[width=\textwidth]{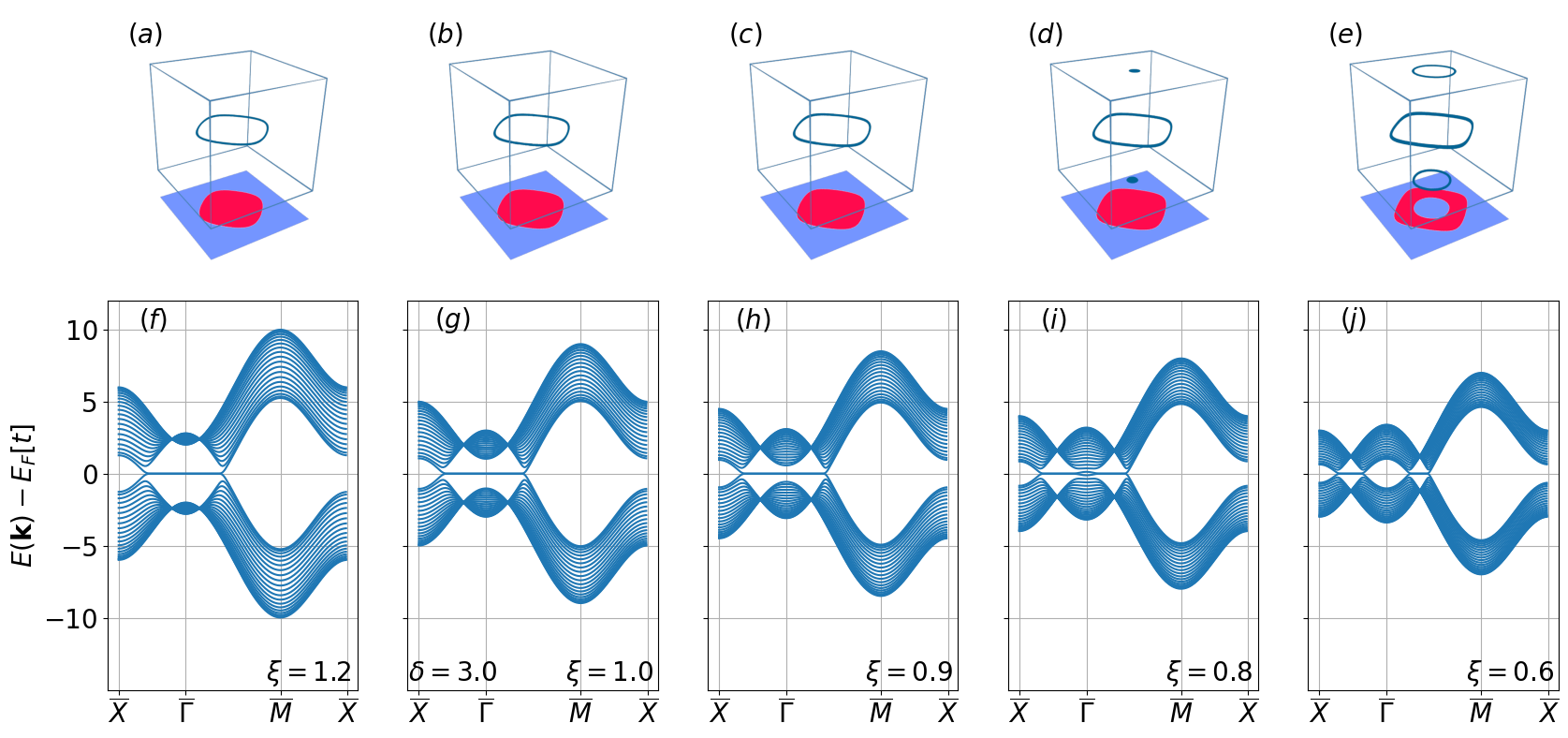}
\caption{\label{fig:band_structure_all_xi}
The geometry of the nodal loop in the three dimensional Brillouin zone, winding number map and slab band structures as above for a fixed value of $\delta=3.0$ and various values of $\xi$.
}
\end{minipage}
\vspace{20mm}
\end{figure*}

\subsection{\label{sec:kspace}Momentum space structure}
As the investigated system is cubic the corresponding Brillouin zone spanned by reciprocal lattice vectors $\mathbf{b}_i$ will also be cubic as depicted in Fig.~\ref{fig:real_and_BZ}(b). As we will connect the topological properties of the bulk to the surface magnetic properties of a slab with finite thickness, it is instructive to introduce the projected Brillouin zone with its appropriate high symmetry points, as shown in the figure, too.
In order to elucidate the momentum space structure defined by the kinetic Hamiltonian \eqref{eq:H0}
we introduce Fourier transformed operators as

\begin{align}
\label{eq:FT}
\hat{a}_{\mathbf{k},s} = \sum_{\mathbf{r}}e^{i\mathbf{k}\mathbf{r}}\hat{a}_{\mathbf{r},s}  \,, \quad
\hat{b}_{\mathbf{k},s} = \sum_{\mathbf{r}}e^{i\mathbf{k}\mathbf{r}}\hat{b}_{\mathbf{r},s},
\end{align}
where $\mathbf{k}$ is a wavevector indexing states in the three dimensional Brillouin zone. With these we can recast \eqref{eq:H0} as
\begin{equation}\label{eq:H0k}
\begin{aligned}
\hat{H}_0 = \sum_{\mathbf{k},s} 
\begin{pmatrix}
\hat{a}_{\mathbf{k},s}^\dagger & \hat{b}_{\mathbf{k},s}^\dagger
\end{pmatrix}
\mathcal{H}(\mathbf{k})
\begin{pmatrix}
\hat{a}_{\mathbf{k},s} \\ \hat{b}_{\mathbf{k},s},
\end{pmatrix}
\end{aligned}
\end{equation}
 where we introduce the matrix $\mathcal{H}(\mathbf{k})$ as 
 \begin{align}\label{eq:mcH}
 \mathcal{H}(\mathbf{k}) &= \left[\delta t_z -2\sum_{i = (x,y,z)}t_i \cos{k_i}\right]\sigma_{x} + 
 2 t_z \sin{k_z}\sigma_{y}\\
 &=\mathbf{d}_{\delta,\xi}(\mathbf{k})\cdot\boldsymbol{\sigma},
 \end{align}
 with $ t_{x,y} = t $, $ t_z = \xi t $ and $\sigma_{x,y}$ are Pauli matrices acting on the sub-lattice space.
The absence of $\sigma_z$ from the above expression is the fingerprint of sublattice symmetry of the model. Three dimensional Hamiltonians with sublattice symmetry can be characterized by winding number \cite{nodal_basics_hirayama2017topological,nodal_basics_PhysRevLett.89.077002,nodal_basics_yang2022quantum} associated to the $\mathbf{d}_{\delta,xi}(\mathbf{k})$ vector for specific paths in momentum space. The system for a given value of $k_x$ and $k_y$ mimics the behaviour of the SSH model \cite{SSH}. We calculate this winding number along $k_z$  as we cross the Brillouin zone. For a given value of $k_x$, $k_y$, $\delta$ and $\xi$ the winding number is evaluated as 
\begin{align}
\nu\left(k_x,k_y,\delta,\xi\right)=\begin{cases}
1 & \left|C_{\delta,\xi}(k_{x},k_{y})/2\xi t\right|<1\\
0 & \left|C_{\delta,\xi}(k_{x},k_{y})/2\xi t\right|>1
\end{cases}
\end{align}
where we introduce the shorthand $C_{\delta,\xi}(k_{x},k_{y})=\delta \xi t -2 t \cos{k_x}-2 t \cos{k_y}$.
The winding number, which is a bulk property, signals the presence or absence of topological drumhead states for slabs. This is a manifestation of the bulk boundary correspondence \cite{asboth2016short}.   
If the winding number is nonzero for a given set of bulk parameters $\delta$ and $\xi$ and wavevector components $k_x$ and $k_y$ then in a slab geometry there will be a zero energy surface state present at the corresponding wavevector. 

The geometry of the nodal loop, the map of winding number and the spectrum of a slab of a finite thickness can be observed for different values of $\delta$ but fixed values of $\xi$ in Fig.~\ref{fig:band_structure_all}. while in Fig.~\ref{fig:band_structure_all_xi}. the same is depicted but for fixed values of $\delta$ and changing $\xi$. 

Let us discuss the evolution of the nodal loop and the drumhead states associated with it as the function of the parameters $\delta$ and $\xi$! 

First, focusing on Fig.~\ref{fig:band_structure_all}. that is keeping $\xi=1.0$, we can observe that, as one decreases $\delta$, a nodal loop first appears at the $\Gamma$ point of the bulk Brillouin zone, then it grows in size. At $\delta=2.0$ two drastic changes occur. First, the nodal loop around the $\Gamma$ point is enlarged to a point where it coalesces with nodal loops from the neighboring Brillouin zone effectively transforming itself from a loop around $\Gamma$ to a loop around $M$. Second, an additional nodal loop is germinated at the $Z$ point of the bulk Brillouin zone, due to a band crossing. The appearance and evolution of the nodal loops leave an impression on the winding number maps as well. For larger values of $\delta$ where only a single loop is present, the region with $\nu=1$ is a simply connected region in the shadow of the nodal loop. For $\delta<2.0$ however, the appearance of the second loop and the coalescence of the original loop causes a drastic change in the connectivity of the region with a finite winding number, changing a simply connected region into a multiply connected one. Let us denote this type of transition as a connectivity shift. This transition is similar to a Lifshitz transition whereby the topology of the Fermi surface changes. 
However, in contrast to the case of other systems with a two-dimensional Brillouin zone, for instance, bilayer graphene \cite{lemonik2010spontaneous,Marcin_Lifshitz_bilayer}, in our special case the Fermi-surface is also a two-dimensional object. 
As $\delta$ is decreased even further to $\delta=0.0$ the area $\Omega_0$ of the region with $\nu=1$ reaches a maximum. Let us introduce the ratio $r$ of this area to the total area of the projected Brillouin zone $\Omega_{BZ}$ as 
\begin{equation}\label{eq:definition_of_r}
    r = \frac{\Omega_0}{\Omega_{BZ}}.
\end{equation}
As expected, due to the bulk boundary correspondence of topological systems, finite winding numbers herald non-dispersing zero energy surface states. As one can observe in Fig.~\ref{fig:band_structure_all}.(f)-(j) where the spectrum of a slab with finite thickness is depicted, the region corresponding to $\nu=1$ indeed harbors drumhead surface states. The spatial localization of these states follows from their analogy with the SSH model \cite{SSH}. 
Turning now our attention to the parameter $\xi$ and to Fig.~\ref{fig:band_structure_all_xi}. we can see that for a fixed value of $\delta$ the parameter $\xi$, which mimics mechanical distortions, can also be used to change the connectivity of the flat portion of the surface localized zero energy states. As one decreases $\xi$ a band crossing can be engineered at the $Z$ point, introducing again a second nodal loop, and thus transforming a simply connected disk like region with $\nu=1$ into an annulus like region. This thus again leads to a connectivity shift.

\subsection{\label{sec:level3}Interactions}
\begin{figure}
    \includegraphics[width=\columnwidth]{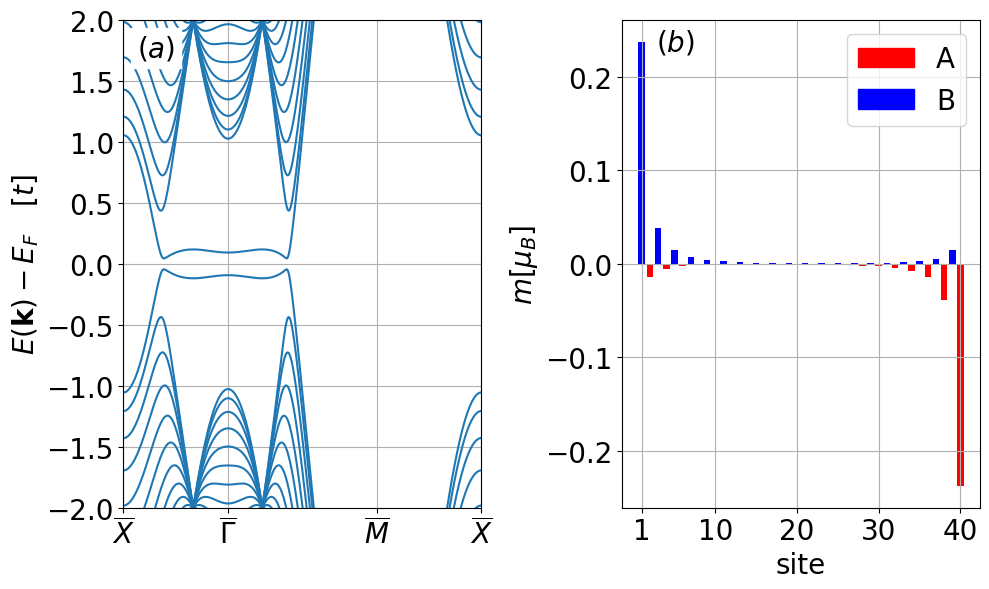}
\caption{Self-consistent  band structure for both spin species (a) of a slab of $N=20$ layers after adding the Hubbard term with $U/t=1.0$ to all sites, for an antiferromagnetic alignment of the top and bottom layers.  Site magnetization $m_i$ (b) of the many-body ground state obtained after the self-consistent procedure.\label{fig:hubbard}}
\end{figure}

In the previous sections, we showed that the presented model exhibits drumhead surface states. For these states, which occupy a considerable portion of the projected Brillouin zone, the kinetic energy vanishes. Interactions between charge carriers thus undoubtedly will have a major role in influencing their behavior. The simplest of consequences of interactions might lead to the formation of an ordered magnetic pattern on the surface of the system. This emergent magnetism parallels that of the edge magnetization of zigzag graphene nanoribbons already observed experimentally \cite{magda2014graphene_ribbon}.

We take interactions into account through a Hubbard term, thus the full Hamiltonian $\hat{H}$ for the electronic degrees of freedom is cast in the form
\begin{equation}\label{eq:Hubbard}
    \hat{H}=\hat{H}_0 + U \sum_{i} \hat{n}_{i ,\uparrow} \hat{n}_{i, \downarrow}, 
\end{equation}
where $\hat{n}_{i,s} = \hat{c}^{\dagger}_{i,s} \hat{c}_{i,s}$ with $\hat{c}_{i,s} = \hat{a}_{\mathbf{r}_i,s}, \hat{b}_{\mathbf{r}_i,s}$. In the present work, we shall focus on the case of a half filled system, thus, in all calculations, the Fermi level $E_F$ is set to guarantee this condition. We have to stress here that in order for magnetism to arise the system needs to be in the vicinity of half filling otherwise the spin polarization of the surface states vanishes, this behavior is expected for nodal line semimetals and it was already observed in rhombohedral graphene \cite{hagymasi2022observation}. However we also note that all mechanisms which make the surface states dispersive, by enhancing its kinetic energy, also will extend the range of the chemical potential at which magnetism can be stabilized.    

In order to further proceed, we analyse the system defined by \eqref{eq:Hubbard} on a mean-field level \cite{claveau2014mean}. That is, we obtain an effective spin dependent single particle description of the system after a self-consistent procedure. Thus instead of the interacting Hamiltonian \eqref{eq:Hubbard} we work with the mean-field Hamiltonian $ \hat{H}^s_{MF}\left (\{n_{i,\uparrow},n_{i,\downarrow}\}\right ) $ for spin channel $s$ which depends explicitly on the self-consistently obtained occupation numbers $n_{i,s}$ at each site. The results of such a mean-field calculation can be observed in Fig.~\ref{fig:hubbard}. (a), where the spectrum of a slab with finite thickness is presented. 
The impact of interactions is the visible splitting of the zero energy flat band. The splitting is due to the local difference of the occupation of the two spin species on the surfaces of the system. The magnetization $m_i$ on site $i$ is obtained as
\begin{equation}
    m_i = (n_{i ,\uparrow} -  n_{i, \downarrow}) \mu_B,
\end{equation}
where the occupation numbers $n_{i,s}$ are the expectation value of $\hat{n}_{i,s}$ in the ground state for site $i$ and spin $s$ while $\mu_{B}$ is the Bohr magneton. 
Fig.~\ref{fig:hubbard}. (b) shows the magnetization for each site in the cross section of a slab of finite thickness. One can observe that the sites on the very top and bottom carry a considerable portion of the overall magnetization. Magnetization drops off exponentially towards the bulk of the system with neighbouring layers exhibiting opposite magnetization.

For moderate system thickness where there is still some overlap between the states localized on the two opposing surfaces of the system, an antiferromagnetic configuration is energetically more favorable where the magnetization of the top layer is reversed as compared to that of the bottom layer, as can be observed in Fig.~\ref{fig:hubbard} (b). In these situations the ground state of the system possesses an overall spectral gap as can be also seen in Fig.~\ref{fig:hubbard} (a).   
For wide enough slabs though, the difference in ground state energy of the parallel and anti-parallel alignment of the magnetization of the opposing surfaces vanishes as the two surfaces effectively decouple from each other.

\section{\label{sec:magnons}Surface magnons}

In this section, we are going to analyze the magnetic characteristics of the topmost surface sites of our model. This layer of sites is characterized at zero temperature by an ordered ferromagnetic spin configuration. We start by mapping the localized magnetic moments of the surface, with magnitude $m$, to that of an isotropic Heisenberg model. The mapping will allow us to find the surface magnon spectrum of the system. From the magnon spectrum, we extract experimentally accessible quantities such as the spin wave stiffness $D$ and the effective exchange constant $J(\mathbf{0})$. We finish this section by discussing how these quantities depend on the parameters of the model. We shall concentrate on possible observable fingerprints of the connectivity shift discussed in the previous sections.  

The classical Heisenberg model describes coupled classical magnetic moments at site $i$ with an orientation $\mathbf{e}_{i}$ and coupling constants $J_{i j}$ through the classical Hamiltonian 
\begin{equation}
h = -\frac{1}{2} \sum_{i, j} J_{i j} \mathbf{e}_{i} \mathbf{e}_{j}.
\end{equation}
For tight binding like electronic systems, with a single spinfull orbital on each site, where interactions are taken into account through a Hubbard term with interaction strength $U$, on the mean-field level, the coupling constants appearing in the above expression can be cast into the rather simple form \cite{oroszlany2019exchange}
\begin{equation}
 J_{i j}= \frac{2}{\pi} \left (\frac{mU}{\mu_B}\right)^2 \sum_{i \neq j}  \int_{-\infty}^{E_{\mathrm{F}}} \mathrm{d} E \operatorname{Im} \left[ {G}^{\uparrow}_{i j}(E) {G}^{\downarrow}_{ j i}(E)\right].
\end{equation}

In this expression $G^s_{ij}(\varepsilon)$ are the matrix elements of the Green's function $\hat{G}^s(\varepsilon)$ for spin channel $s$ and between surface sites $i$ and $j$ which in turn are obtained from the mean-field Hamiltonian $\hat{H}^s_{MF}$ as 
\begin{equation}
    \hat{G}^s(E)=\lim_{\eta\rightarrow 0}\left ((E+i\eta) \hat{I}-\hat{H}^s_{MF}\right)^{-1}.
\end{equation}
The Fourier transform of the coupling constants, $J(\mathbf{q})$, can be cast in terms of an integral over the projected Brillouin zone for each wave vector $\mathbf{q}$ as
\begin{equation}
J(\mathbf{q})=\sum_{j \neq 0} e^{i \mathbf{q} \mathbf{R}_{j}} J_{0 j} 
= \frac{2}{\pi} \left (\frac{mU}{\mu_B}\right)^2  \operatorname{Im} \int_{-\infty}^{E_{\mathrm{F}}} \mathrm{d} \varepsilon \mathcal{I}_\mathbf{q} (E )
\end{equation}
with
\begin{equation}
    \mathcal{I}_\mathbf{q} (E ) =  \left(\sum_{\boldsymbol{k}} \mathcal{G}^{\uparrow}_{00}(E, \mathbf{k}) {\mathcal{G}}^{\downarrow}_{00}(E, \mathbf{k}+\mathbf{q}) \right) - {G}_{00}^{\uparrow}(E) {G}_{00}^{\downarrow}(E).
\end{equation}
Here ${\mathcal{G}}^{s}_{00}(E, \mathbf{k})$ is the surface component of the momentum dependent Green's function for an infinite slab geometry of finite thickness at momentum $\mathbf{k}$ and spin component $s$. 
The coupling constants can be used to define a temperature scale analogous to the mean-field Curie temperature as $J(\mathbf{0})/3k_B$. Thus we shall use $J(\mathbf{0})$, the effective exchange parameter \cite{liechtenstein1987local}, as a key characteristic property as well.  

The dynamics of spin fluctuations is captured by the dispersion relation of magnons, which in turn, for a ferromagnetic reference state, is given by
\begin{equation}
    \varepsilon(\mathbf{q})=\frac{2 \mu_{B}}{m}(J(\mathbf{0})-J(\mathbf{q})).
\end{equation}
This spectrum can be measured for instance by spin polarized electron energy loss spectroscopy\cite{zakeri2013surface_magnon_SPEELS,zakeri2021unconventional_surface_magnon_SPEELS}.
For ferromagnetic systems the curvature $D$ of the magnon spectrum at $\mathbf{q}=\mathbf{0}$ is again an important attribute which is more commonly referred to as spin wave stiffness. That is 
\begin{equation}
 \varepsilon(\mathbf{q})|_{\mathbf{q}\approx\mathbf{0}} = D {q} ^2.
\end{equation}

In the following we present and discuss results for the quantities mentioned above. We put an emphasis on how the energetics of surface magnons are impacted by the two model parameters $\delta$ and $\xi$ particularly around a connectivity shift of the surface flat band. 
Finite size scaling shows that as one increases the number of layers $N$ towards the macroscopic limit the identified signatures of the connectivity shift presented below will manifest precisely at the critical values of parameters, even for weak interaction strengths. For stronger interactions the fingerprints of the transition will occur already for a moderate number of layers. In the calculations shown we considered a slab of thickness $N=20$ layers and an interaction strength of $U/t=1.0$ which proved to be a pragmatic choice in order to illustrate our main message.

\begin{figure}
\includegraphics[width=\columnwidth]{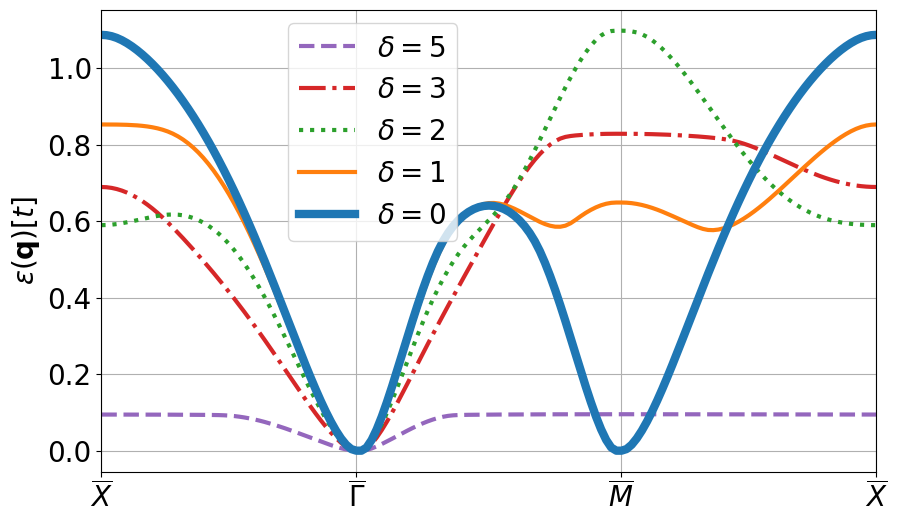}
\caption{Surface magnon spectrum of a slab for different values of $\delta$ with $\xi=1.0$. \label{fig:magnon_for_delta}}
\end{figure}
\begin{figure}
\includegraphics[width=\columnwidth]{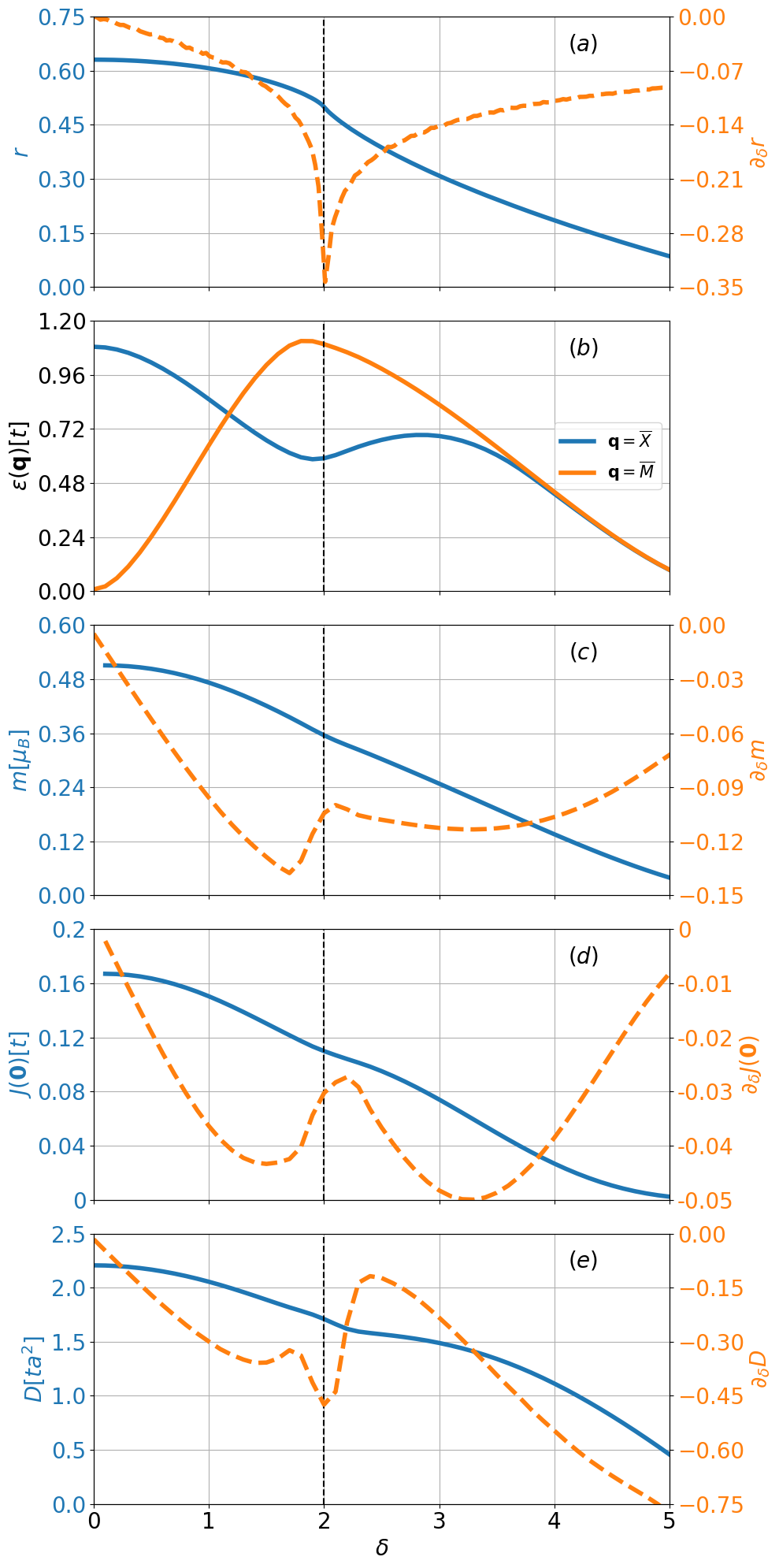}
\caption{The ratio $r$ and its derivative with respect to $\delta$ (a), the
surface magnon spectrum at high symmetry points (b), magnetization $m$ and its derivative (c), effective exchange parameter $J(\mathbf{0})$ with its derivative (d) and stiffness constant $D$ with its derivative (e) for different values of $\delta$ at $\xi=1.0$. Dashed vertical line at $\delta=2.0$ signals the connectivity shift. \label{fig:props_delta}}
\end{figure}
As we did in previous sections we start our analysis by focusing on the parameter $\delta$ and keeping $\xi=1.0$ that is we consider a system in the absence of mechanical distortions. The magnon spectrum around a high symmetry path of the projected Brillouin zone for various values of $\delta$ is depicted in Fig.~\ref{fig:magnon_for_delta}. As one can deduce from the graph reducing the value of $\delta$, increases the energy of magnons around the $\overline{\Gamma}$ point. A curious observation can be also made regarding the spectrum for $\delta=0.0$, namely that it vanishes not just at $\overline{\Gamma}$ but also at $\overline{M}$. This property, which would point towards the instability of the ferromagnetic phase in general, can be explained in this particular case. In this instance the absence of the hopping terms proportional to $\delta$ from the kinetic term $\hat{H}_0$ means that the system falls apart in to two interlocked but decoupled subsystems, which can be oriented parallel or anti-parallel with respect to each other without any energy cost. 
In order to further elucidate important characteristic features of the obtained magnon spectrum we plot key properties as a function of $\delta$ in Fig.~\ref{fig:props_delta}. We comment first on the evolution of $r$ depicted in subfigure (a). As the nodal loop enlarges with decreasing $\delta$ the drum-head surface states occupy more and more area from the projected Brillouin zone. However decreasing $\delta$ beyond the connectivity shift at $\delta=2.0$ the growth of the ratio $r$, depicted by orange dashed line in the figure, suffers a discontinuity. A qualitative observation regarding the connectivity shift can be also made based on the evolution of the magnon energies at the high symmetry points shown in subfigure (b). A maximum in the vicinity of the connectivity shift at the $\overline{M}$ point while a local minimum at the $\overline{M}$ point is present. Signatures of the connectivity shift are also present in the magnetization $m$ the effective exchange coupling $J(\mathbf{0})$ and in the stiffness $D$ visualized in subfigures (c),(d) and (e) respectively. Although somewhat hard to discern these directly they are more readily visible through their derivatives with respect to $\delta$. The derivative of the magnetization $\partial_\delta m$ jumps, the derivative of the effective coupling  $\partial_\delta J(\mathbf{0})$ shows a local maximum while the derivative of the stiffness $\partial_\delta D$ has a local minimum in the vicinity of the connectivity shift at $\delta=2.0$. 

\begin{figure}
\includegraphics[width=\columnwidth]{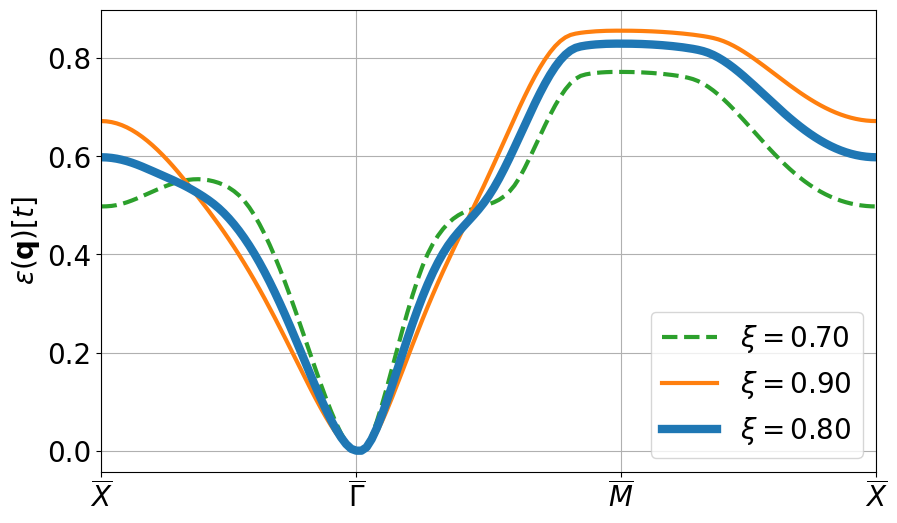}
\caption{Surface magnon spectrum of a slab for different values of $\xi$ with $\delta=3.0$. \label{fig:magnon_for_xi}}
\end{figure}
\begin{figure}
\includegraphics[width=\columnwidth]{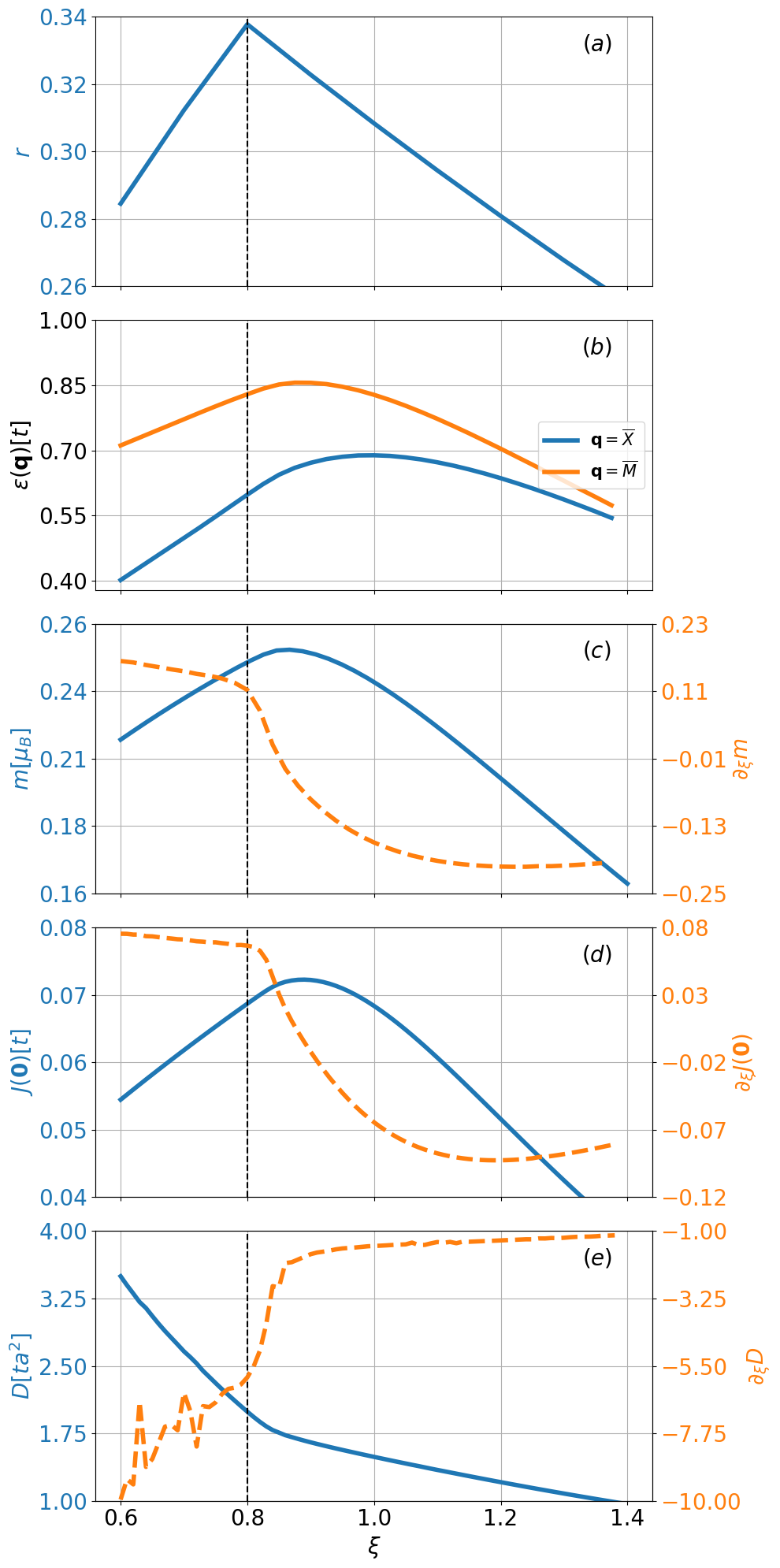}
\caption{The ratio $r$ (a), surface magnon spectrum at high symmetry points (b), magnetization $m$ (c), effective exchange parameter $J(\mathbf{0})$ (d) and stiffness constant $D$ (e) for different values of $\xi$ with $\delta=3.0$. Where appropriate the right axis shows the scale of the derivative with respect to $\xi$. The vertical dashed line at $\xi=0.8$ marks the connectivity shift.\label{fig:props_xi}}
\end{figure}
In an experimental setting the parameter $\delta$ is typically hard to control, $\xi$ on the other hand is directly linked to a uniaxial distortion of the sample  in the $z$ direction. As we discussed previously a connectivity shift occurs for $\delta=3.0$ if we decrease $\xi$ below the critical $0.8$ value, thus examining the behaviour of the above detailed characteristic features for this case as well might highlight experimentally observable  fingerprints of this transition.
In Fig.~\ref{fig:magnon_for_xi}. the magnon dispersion relation is depicted for distinct values of $\xi$ above, below and exactly at the connectivity shift. In the panels of Fig.~\ref{fig:props_xi}. the detailed $\xi$ dependence of the characteristic magnon spectral features is collected. The discontinuity of the evolution of the ratio $r$ at the connectivity shift is evident as in this case $r$ peaks at the transition point. The magnon energies at the high symmetry $\overline{M}$ and $\overline{X}$ point as well as the magnetisation and the effective exchange coupling show a local maximum in the vicinity of the connectivity shift, while in the evolution of the stiffness a considerable decrease in the slope is observable as $\xi$ increases past the transition point. 
In this case it will be also insightful to evaluate the derivatives, now with respect of $\xi$. In all the characteristic properties there is a clear transition happening at the connectivity shift in the derivatives. The derivative of $m$ and $J(\mathbf{0})$ both drop sharply while the $\partial_\xi D$ jumps abruptly at the transition point. We note that the oscillations present in this quantity at small $\xi$ values are due to numerical limitations, and as such they should be considered as a computational artefact.

\section{Summary}
In conclusion, we investigated the magnons associated to the drumhead surface states in a simple model of a nodal loop semimetal.
The model without interactions exhibits topological flat bands whose shape, and crucially their connectivity, can be controlled by mechanical distortions. 
Including interactions on a mean-field level we show that magnetization on the surface is stabilized.
Employing a standard Green's function based technique we obtained the dispersion relation of surface magnons.
Determining key, experimentally accessible characteristics of the magnon spectrum, such as the magnetization, the effective exchange coupling and the spin wave stiffness, we show that the Lifshitz-like transition of the electronic states can in principle be observed through the magnetic properties of the surface.

On the one hand we emphasise that our presented phenomenological observations would greatly benefit from future analytic calculations which may shed light to the intricate interplay of topology, interactions and magnetism in this system. 
On the other hand our calculations hopefully will encourage experimental exploration of magnetism on the surface of nodal loop semimetals. For instance  Ca$_3$P$_2$ \cite{ca3p2_experiment} with a relative large $r$ ratio might be an excellent candidate for future investigations.

\section{Acknowledgement}
The authors wish to express their gratitude to Edward McCann, Rahul Nandkishore, Jaime Ferrer, Amador García Fuente, Gabriel Martinez Carracedo, László Szunyogh and László Udvardi, for valuable discussions and their comments regarding the present work.
This research was supported by the Ministry of Culture and Innovation and the National Research, Development and Innovation Office within the Quantum Information National Laboratory of Hungary (Grant No. 2022-2.1.1-NL-2022-00004) and by NKFIH Grants No. K131938, K134437 and K142179.  A.A. greatly acknowledges the support from Stipendium Hungaricum No. 249316. 
L.O. also acknowledges support of the National Research, Development and Innovation (NRDI) Office of Hungary and the Hungarian Academy of Sciences through the Bolyai and Bolyai+ scholarships. 

\newpage
\bibliography{refs}

\end{document}